# Three attempts at two axioms for quantum mechanics

**Daniel Rohrlich**
Physics Department, Ben Gurion University of the Negev, Beersheba
rohrlich@bgu.ac.il

**Abstract:** The axioms of nonrelativistic quantum mechanics lack clear physical meaning. In particular, they say nothing about nonlocality. Yet quantum mechanics is not only nonlocal, it is twice nonlocal: there are nonlocal quantum correlations, and there is the Aharonov-Bohm effect, which implies that an electric or magnetic field *here* may act on an electron *there*. Can we invert the logical hierarchy? That is, can we adopt nonlocality as an axiom for quantum mechanics and *derive* quantum mechanics from this axiom and an additional axiom of causality? Three versions of these two axioms lead to three different theories, characterized by "maximal nonlocal correlations", "jamming" and "modular energy". Where is quantum mechanics in these theories?

Among Itamar Pitowsky's many admirable qualities, I admired most his capacity for thoroughly exploring incompatible points of view, approaches and theoretical frameworks. We tend to ignore approaches that are incompatible with our own. It is a natural tendency. It takes work to overcome it. Itamar worked hard to understand all points of view, which led to another of his admirable qualities, his comprehensive knowledge. He was a true *philosopher* – love of knowledge and understanding animated him. As a result, whether in a seminar on campus or at a demonstration on a street corner, he most often understood all the points of view better than anyone else did. The Talmud (ברכות סד, א) expresses this capacity in a adage, "תלמידי חכמים מרבים שלום בעולם" – "scholars bring peace to the world" – which Rabbi Avraham Kook (1865–1935) explained as follows:

> Some people mistakenly believe that world peace demands uniformity of views and practices. When they see scholars studying philosophy and Torah and arriving at a plurality of views and approaches, they see only controversy and discord. But – on the contrary – true peace comes to the world only by virtue of its plurality. The plurality of peace means appreciating each view and approach and seeing how each has its own place, consistent with its value, context and content.

(From עולת ראיה ח"א ע' של; translations from Hebrew are mine.) One theoretical framework that Itamar explored [1] is the one that I have the honor to present here.

●

Quantum mechanics doesn't supply its own interpretation. The numerous competing interpretations of quantum mechanics testify to the fact that quantum mechanics doesn't supply its own interpretation. About one of these interpretations, due to H. Everett [2] and J. A. Wheeler [3], Bryce DeWitt said the following:

> Everett took the quantum theory the way it was and didn't impose anything on it from the outside. No classical realm. Does the theory produce its own interpretation? If so, how? And he, in my view he's the only one who's answered properly in the affirmative that it provides its own interpretation [4].

Yet, more than half a century after Everett's bold and provocative paper, the Everett-Wheeler interpretation remains exactly that – one more competing interpretation, with no consensus in sight. This irony, as well, testifies to the fact that quantum mechanics doesn't supply its own interpretation. What would it take to show that, on the contrary, quantum mechanics does supply its own interpretation? It would take an interpretation so natural and satisfactory as to induce consensus. But how could such an interpretation ever spring from the opaque axioms of quantum mechanics? If we seek a theory that has clear, unambiguous physical meaning, let us derive it from axioms that have clear, unambiguous physical meaning. This paper argues that the way towards a natural and satisfactory interpretation of quantum mechanics passes through new and physically meaningful axioms for the theory, and reports attempts to define such axioms.

The axioms of quantum mechanics are notoriously opaque. Without trying or needing to be comprehensive, let us mention a few of them. "Any possible physical state corresponds to a ray in a Hilbert space." "Physical observables correspond to Hermitian operators." Do these axioms tell us something about the physical world? Or do they merely list mathematical structures useful for describing the world? If they merely list useful mathematical structures, what is it about the underlying physics that makes these structures useful? What we want, after all, is the physics. "If $P(a,\psi)$ is the probability that a measurement of a physical observable $A$, corresponding to a Hermitian operator $\hat{A}$, on a system in the normalized state $|\psi>$, yields $a$, then $P(a,\psi) = <\psi| \Pi_a |\psi>$, where $\Pi_a$ projects onto the subspace of eigenstates of $\hat{A}$ having eigenvalue $a$." This axiom, too, is opaque; it offers no hint as to whether the probability $P(a,\psi)$ is intrinsic or derives from an underlying determinism. It is more of an algorithm than an axiom.

Long ago, Yakir Aharonov [5] drew an analogy that is at once amusing and penetrating. The special theory of relativity, we know, follows elegantly from two axioms: first, the laws of physics are the same for observers in all inertial reference frames; second, the speed of light is a constant of nature. These axioms have clear physical meanings: the first specifies a fundamental space-time symmetry and the second specifies a physical constant. Suppose we had, instead of these two standard axioms, three "nonstandard" axioms: first, physical objects contract in the direction of their motion (FitzGerald contraction); second, this contraction and a "local time" (lacking clear physical significance) combine in "Lorentz transformations" that form a group; third, simultaneity is subjective. Even supposing we could deduce the special theory of relativity from these three axioms, would there be consensus about its interpretation? What kind of consensus *could* there be, as long as we were stuck with the wrong axioms? Analogously, perhaps consensus about how to interpret quantum mechanics is elusive because we are stuck with the wrong axioms for the theory. Perhaps we have grasped quantum mechanics by the tail, or by the hind legs, instead of by the horns.

If this analogy seems strained, let us note that all three of these "nonstandard" axioms had proponents (not including A. Einstein) already in 1905. Moreover, as late as 1909, H. Poincaré delivered a series of lectures in Göttingen, culminating in a lecture on "La mécanique nouvelle". At the basis of Poincaré's "new mechanics" were three axioms, the third being the FitzGerald contraction: "One needs to make still a third hypothesis, much more surprising, much more difficult to accept, one which is of much hindrance to what we are currently used to. A body in translational motion suffers a deformation in the direction in which it is displaced." [6] In Poincaré's new mechanics, the FitzGerald contraction was an axiom in itself, not a consequence of other axioms, hence Poincaré found it "much more difficult to accept" than we do. Well, aren't all the axioms of quantum mechanics surprising, difficult to accept, and of *much hindrance* to what we are used to?

We can take the analogy further [5]. The logical structure of the special theory of relativity is exemplary. From its two axioms we can deduce all the kinematics of the theory, and with scarcely more input we can deduce all the dynamics as well. How can two axioms be so efficient? If we look at the axioms with a Newtonian mindset, we see clearly that the two axioms contradict each other. Well, of course they don't contradict each other. But they come so close to contradicting each other, that a unique mechanics reconciles them: Einstein's special theory of relativity. Now in the quantum world, as well, we have two physical principles that come close to contradicting each other. One is the principle of causality: *relativistic* causality, also called "no signalling", is the principle that no signal can travel faster than light. The other principle is nonlocality. Quantum mechanics is nonlocal, indeed twice nonlocal: it is nonlocal in two inequivalent ways. There is the nonlocality of the Aharonov-Bohm [7] and related [8] effects, and there is the nonlocality implicit in quantum correlations that violate Bell's [9] inequality. Let us briefly comment on each of these.

One often reads that the Aharonov-Bohm effect proves that the scalar and vector potentials of electromagnetism have a degree of reality in quantum physics that they do not have in classical physics. This distinction is valid, but it can mislead us into thinking that the Aharonov-Bohm effect is local, because an electron diffracting around a shielded solenoid or capacitor interacts locally with the vector potential of the magnetic field in the solenoid, or with the scalar potential of the electric field in the capacitor. However, there is no way to measure these potentials. What is measurable are the magnetic and electric fields. The dependence of the electron diffraction pattern on these fields, i.e. the Aharonov-Bohm effect, must be nonlocal, because the electron does not pass through them. Yet the fields are produced locally; so one might try (i.e. by adjusting the capacitor or the current in the solenoid) to send a superluminal signal that would show up in the electron diffraction pattern. It is indeed possible to send signals that show up in the interference pattern, but they are never superluminal.

Bell's inequality implies that quantum correlations can be nonlocal in the following sense. Let Alice and Bob, shown in Fig. 1, be two experimental physicists,

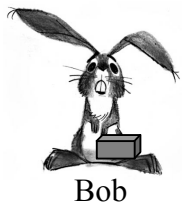
Bob

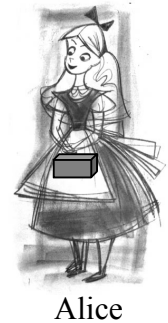
Alice

Fig. 1. Alice and Bob (drawn by Tom Oreb © Walt Disney Co) with their respective black boxes.

in their respective labs. Note that each is equipped with a black box. From spacelike-separated measurements on their respective black boxes, Alice and Bob obtain correlations that violate Bell's inequality. They may suppose that the correlations already existed (as local "plans" or programs) in the black boxes before their measurements, but this explanation is what Bell ruled out. Another explanation – that the black boxes send superluminal signals to each other – might suggest to them that they, too, could send superluminal signals to each other. But they cannot [10]. Quantum correlations can violate Bell's inequality but they are useless for signalling.

The coexistence of nonlocality and causality in the Aharonov-Bohm effect and in nonlocal quantum correlations is remarkable. If there are nonlocal effects, what stops us from signalling with them? Quantum mechanics must be an extraordinary theory if it can do this trick – if it can make nonlocality and causality coexist. A. Shimony [11-12] speculated that quantum mechanics might be the only theory that can do so, i.e. that quantum mechanics might follow uniquely from the two axioms of nonlocality and causality. Aharonov [5] even suggested a part of the logical structure. If we look at the details of just how nonlocality coexists with causality, we discover they always involve quantum uncertainty. Indeed, quantum mechanics has to be a probabilistic theory, because if nonlocal influences were certain, how could they *not* violate causality? Aharonov's suggestion thus inverts the conventional logical hierarchy of quantum mechanics: instead of making probability an axiom and deriving nonlocality as a quantum effect, he makes nonlocality an axiom and derives probability (uncertainty) as a quantum effect.

Inspired by Aharonov's suggestion, I will now describe three attempts [13-15] to define more precisely the axioms of "causality" and "nonlocality" and to derive quantum mechanics from them. The three attempts fall under the headings "Maximally nonlocal correlations" (or "PR boxes"), "Jamming", and "Modular energy".

**I. Maximally nonlocal correlations**

Our goal is to derive quantum mechanics, but we have still to decide whether we should try to derive nonrelativistic or relativistic quantum mechanics. We have two reasons to decide for nonrelativistic quantum mechanics. First, it is a simpler theory with a superior formalism. Relativistic quantum mechanics naturally allows creation of particle-antiparticle pairs; it is more complicated in that the number of particles is not fixed. Even if we artificially fix the number of particles, the formalism of relativistic quantum mechanics is not satisfactory: not all of its Hermitian operators correspond to physical observables [16]. That is, not all of its Hermitian operators are measurable in practice. Relativistic causality imposes constraints on what is measurable in practice. However, these constraints do not apply to the Hermitian operators of nonrelativistic quantum mechanics; they are all measurable [17]. If our goal is to find the right axioms for quantum mechanics, we have a better chance of finding them for a theory that already has a satisfactory formalism. It is also plausible that if we find the right axioms for the nonrelativistic theory, we will be in a better position to find the right axioms for the relativistic theory, axioms that do not lead to unmeasurable Hermitian operators.

The second reason to decide for nonrelativistic quantum mechanics has to do with the axiom of nonrelativistic causality. According to this axiom, no signal can travel faster than the speed of light, $c$. In the nonrelativistic limit, $c$ is infinite, so the axiom of relativistic causality appears, initially, to become vacuous. But closer inspection reveals that the axiom of relativistic causality is actually *stronger* in this limit: it tells us that quantum correlations cannot be used to send signals at *any* speed. How so? Suppose that the quantum correlations measured by Alice and Bob could be used to send a signal. These correlations do not depend on whether the interval between Alice's and Bob's respective measurements is spacelike or timelike; indeed, nothing in nonrelativistic quantum mechanics can distinguish between spacelike and timelike intervals, because there is no *c*. Hence if quantum correlations could be used to send any signal, they could be used to send a superluminal signal. Hence they cannot be used to send any signal. If our choice of an axiom of nonlocality has to do with nonlocal correlations, the axiom of causality implies that quantum correlations are useless for sending signals at any speed. We find ourselves with two axioms with clear physical meaning:

      1. No quantum correlations can be used to send signals.

      2. Some quantum correlations are nonlocal.

The second axiom means that some quantum correlations violate some version of Bell's inequality. The most suitable version of Bell's inequality for the case of bipartite correlations (measured by Alice and Bob) is due to Clauser, Horne, Shimony and Holt (CHSH) [18]. Let us assume that Alice and Bob make a series of joint measurements on their respective boxes, in which Alice measures either *A* or *A'* and Bob measures either *B* or *B'* in each joint measurement, where *A*, *A'*, *B* and *B'* are physical observables. In each joint measurement they must each choose which observable to measure – they can never measure both – and the results of their measurements are always 1 or –1. From the results of their joint measurements, they can compute four types of correlations: C(*A*,*B*), C(*A'*,*B*), C(*A*,*B'*) and C(*A'*,*B'*), where C(*A*,*B*) is the correlation between the results of their joint measurements when Alice chooses to measure *A* and Bob chooses to measure *B*, and so on. The CHSH inequality states that if these correlations already existed (as local plans or programs) in the black boxes before their measurements, then a certain combination of them is bounded in absolute magnitude:

$$| C(A,B) + C(A,B') + C(A',B) - C(A',B') | \leq 2 \quad . \tag{1}$$

So if the correlations that Alice and Bob measure satisfy this inequality, they are local correlations. Conversely, if the correlations that they measure do not satisfy the inequality, they are nonlocal correlations. Axiom 2 above, the nonlocality axiom, states that at least for some physical variables *A*, *A'*, *B* and *B'* and some preparation of their black boxes, the correlations they measure violate the CHSH inequality.

      Besides the bound 2 of the CHSH inequality, two other numbers, $2\sqrt{2}$ and 4, are important bounds for the left-hand side of Eq. 1. If the correlations C(*A*,*B*), C(*A'*,*B*), C(*A*,*B'*) and C(*A'*,*B'*) were completely independent, the absolute value on the left-hand side could be as large as 4. But if they are quantum correlations, the absolute value on the left-hand side can only be as large as "Tsirelson's bound" [19], which is $2\sqrt{2}$. It can be larger than 2 just as $2\sqrt{2}$ is larger than 2, but it cannot be larger than $2\sqrt{2}$. Now the fact that quantum correlations cannot violate Tsirelson's bound is curious. If they are strong enough to violate the CHSH inequality, why aren't they strong enough to violate Tsirelson's bound?

A plausible answer to this question, in the spirit of Aharonov's suggestion, is that if quantum correlations were any stronger, they would violate causality as well as Tsirelson's bound. Indeed, if quantum mechanics follows from our two axioms, then so does Tsirelson's bound – a theorem of quantum mechanics. Conversely, if Tsirelson's bound does not follow from our two axioms, then certainly quantum mechanics does not. So if we can find "maximally nonlocal" correlation functions $C_{max}(A,B)$, $C_{max}(A',B)$, $C_{max}(A,B')$ and $C_{max}(A',B')$ that obey the two axioms but violate Tsirelson's bound, we must conclude that quantum mechanics does not follow from our two axioms.

It is straightforward to define such correlations. We do it in two steps:

1. In any measurement of $A$, $A'$, $B$ or $B'$, let the results 1 and −1 be equally likely.

2. Let $C_{max}(A,B) = C_{max}(A,B') = C_{max}(A',B) = 1 = -C_{max}(A',B')$.

The first step insures that whatever Alice chooses to measure will not change the probabilities of ±1 as results of Bob's measurement – whether he measures $B$ or $B'$ – and vice versa, because in any case the probabilities of all the results equal 1/2. Hence Alice cannot send a signal to Bob by her choice of what to measure, and vice versa. These correlations satisfy the axiom of causality. At the same time, they are so strong – joint measurements of $A$ and $B$, of $A'$ and $B$, and of $A$ and $B'$ are perfectly correlated, while joint measurements of $A'$ and $B'$ are perfectly anticorrelated – that the left-hand side of Eq. 1 violates the CHSH inequality and Tsirelson's bound maximally.

Our attempt to derive nonrelativistic quantum mechanics from two axioms is apparently a failure, but not necessarily a total failure. We may still hope that, with an additional axiom, we will be able to derive it. Indeed, W. van Dam [20] showed that in a world containing maximally nonlocal correlations, an important class of communication tasks would become dramatically simpler. G. Brassard et al. [21] extended this result to nonlocal correlations that are not maximal, indeed not much stronger than nonlocal quantum correlations. More recently, M. Pawłowski et al. [22] have defined an axiom of "information causality" and shown that any nonlocal correlation violating Tsirelson's bound is incompatible with this axiom. Their results are striking, but the physical meaning of "information causality" is perhaps not sufficiently clear. Maximal nonlocal correlations have even inspired experimental work [23].

## II. Jamming

There is action at a distance in the Aharonov-Bohm effect. A solenoid acts at a distance on a beam of electrons; the interference pattern of the electrons depends on how the experimental physicist prepares the solenoid. By contrast, there is no action at a distance in quantum correlations. If there were, Alice and Bob could use them to send signals to each other; but in quantum correlations there is only "passion at a distance", as Shimony [11] aptly put it. If we define nonlocality as action at a distance, can we derive quantum mechanics (or some part of it) from nonlocality and causality? We shall see that "jamming", a presumably non-quantum form of action at a distance, could coexist with causality. We shall also discuss an axiom for "modular energy", a quantum form of action at a distance.

Having redefined the axiom of nonlocality, we must now redefine also the axiom of causality. We cannot say, "No action at a distance can be used to send signals," when action at distance is itself a signal. We can only say that the signal cannot be outside the forward light cone of the act of sending it. So we are back to relativistic causality. Our two axioms are then

1. There is no superluminal signalling.

2. There is action at a distance.

In the Aharonov-Bohm effect, the act of preparing the solenoid or capacitor produces electromagnetic radiation, which cannot have an effect outside the forward light cone of the preparation. If the effect shows up in the interference pattern, it is always within the forward light cone of the preparation and there is no superluminal signalling.

We return to Alice and Bob, and welcome their friend, Jim the Jammer (in Fig. 2). Jim has a large black box of his own, which acts nonlocally on their boxes. If Jim presses the button on his box, it turns their nonlocal correlations into local correlations, i.e. into correlations that obey the CHSH inequality. The probability of each result 1 or –1 does not change – it remains 1/2 – but the correlations change. Let's assume that their results

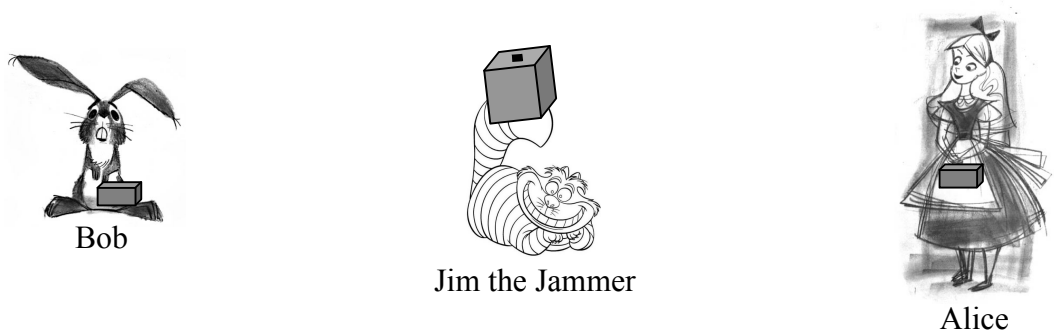

Fig. 2. Alice and Bob, with Jim the Jammer.

become completely uncorrelated. So here we have Jim acting at a distance to change the correlations between the black boxes of Alice and Bob. Does this action at a distance obey the no-signalling constraint? Clearly, Jim cannot send a signal to *either* Alice *or*

Bob, because his action does not change the probabilities of 1 and −1 as results of their measurements. But can Jim send a signal to Alice *and* Bob? Indeed he can, because when Alice and Bob compare their results and compute correlations, they immediately identify their correlations as local or nonlocal. If Alice, Bob and Jim have arranged in advance that local correlations mean "Yes", and nonlocal correlations mean "No" (in some context), then Jim can signal "Yes" or "No" by choosing to jam, or not to jam, their correlations.

Figures 3(a) and 3(b) are spacetime diagrams of jamming (with time on the vertical axis). For simplicity, we reduce jamming to three spacetime events. At $j$, Jim may press the button on his black box. At $a$ and $b$, Alice and Bob make their respective measurements. Now, Alice and Bob can compare their results anywhere, and only, in the overlap of their future light cones. Hence, they can receive Jim's signal anywhere, and only, in the overlap of their future light cones. However, if this overlap does not lie entirely within the future light cone of $j$, Jim's signal can be superluminal. So for jamming to be consistent with the no-signalling axiom, jamming must work *only* when the overlap of the future light cones of $a$ and $b$ lies within the future light cone of $j$.

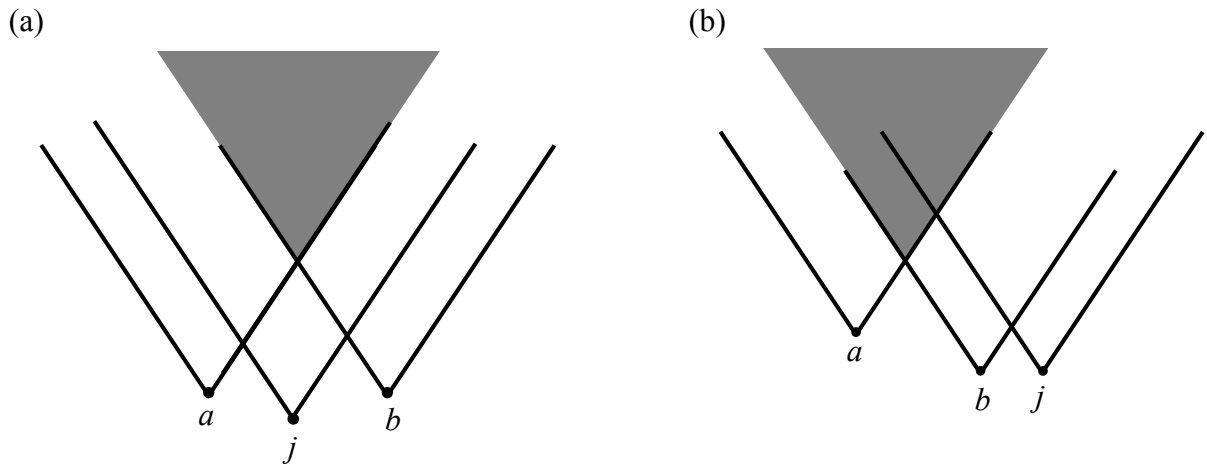

Fig. 3. The overlap of the future light cones of $a$ and $b$ either (a) lies or (b) does not lie entirely within the future light cone of $j$.

Hence in Fig. 3(a) Jim *can* jam their correlations, in Fig 3(b) he *cannot*. Then jamming is consistent with our two axioms.

On the face of it, at least, there is nothing like jamming in quantum mechanics. If not, then we have once again shown that causality and nonlocality – now in the guise of action at a distance – do not imply quantum mechanics as a unique theory: at least one other theory, however rudimentary, is consistent with these two axioms. Another failure!

## III. Modular energy

Twice, starting with a general axiom of nonlocality, we have failed to derive quantum mechanics. What if we start with an axiom of nonlocality that is tailored to quantum mechanics? Can we then derive quantum mechanics (or part of it)? If we succeed, our success will be less impressive because of the initial tailoring; but at least we may succeed. I will now define an axiom of nonlocality that is, in fact, a dynamical form of quantum nonlocality [24]. The axiom of causality, as well, will be tailored to nonrelativistic quantum mechanics.

We begin with the axiom of causality. The speed of light has no special status in nonrelativistic quantum mechanics. For nonlocal correlations, this fact implies that correlations are useless for any kind of signalling, not just superluminal signalling. Our first try at an axiom of causality was, therefore, that no correlations can be used for signalling. But here we will define nonlocal dynamics instead of nonlocal correlations. Our axiom cannot be that no *dynamics* can be used for signalling; causal relations, including signalling, are inherent in dynamics. If Alice sends a particle to Bob, the particle carries a signal. A more plausible axiom of causality is that signalling is possible only via a material interaction, i.e. only when a particle or other object connects the cause and effect. If Alice's particle mysteriously disappears and then reappears in Bob's lab, *that* particle cannot carry a signal, according to this axiom of causality.

We now define an axiom of nonlocality for a specific physical setting that includes Alice and Bob. Fig. 4 shows a shaft with a piston in it at Bob's end. The piston can slide without friction in the shaft. Attached to the outer end of the piston is a box with two open sides. Bob throws a ball that ricochets through the box in two elastic collisions, pushing the (stationary) piston in a short distance. Alice's end of the shaft is closed, yet she can release a particle at her end. Our axiom of nonlocality states that the ball and particle – *if* Alice releases a particle – exchange energy *nonlocally* in the time between the two collisions.  Energy is conserved, but it can mysteriously disappear from Alice's lab and reappear in Bob's, even before the particle reaches the piston.

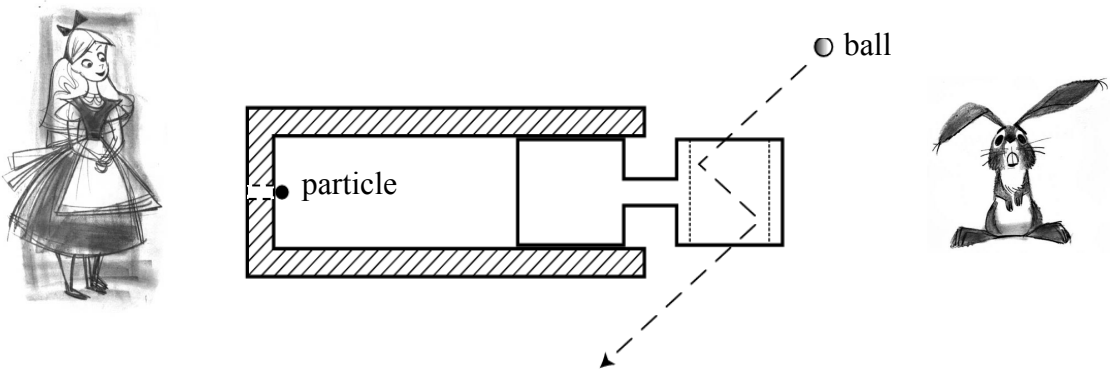

Fig.4. Alice may release a particle at her end while Bob throws a ball at his end.

On the face of it, this axiom contradicts the causality axiom.  If Alice chooses to release a particle into the shaft, Bob will detect a change in the energy of the ball; if she chooses not to, he will detect no change.  Thus Alice can signal to Bob via her choice whether or not to release a particle.  In the spirit of Aharonov's suggestion, we can infer that Bob's measurements of energy must be uncertain.  Even so, there seems to be no way to reconcile the nonlocality axiom with the causality axiom.  Let the energies $E_A$ and $E_B$ of the particle and the ball, respectively, be distributed according to probability distributions, so that the energy of each is uncertain.  If the particle enters the shaft, we expect the probability distribution of $E_B$ to change.  If so, Alice and Bob can still use nonlocal energy exchange to send a signal – if not in one run of the experiment, then in many simultaneous runs on many copies of the experiment:  Alice's signal will emerge from the statistics of Bob's measurements.

Yet mathematical analysis [24] reveals an additional possibility.  Let $E_0$ be a parameter with units of energy.  For any $E_0$ and for any energy $E$, we define an associated quantity $E_{mod} = E$ modulo $E_0$ (also written $E_{mod} = E$ mod $E_0$), which is the energy $E$ minus a multiple of $E_0$ such that $0 \leq E_{mod} < E_0$.  We call this quantity *modular* energy.  If the particle and the ball exchange *modular* energy, they exchange at most $E_0$ in energy. Since the total energy $E_A + E_B$ is conserved in any exchange, so is the total modular energy $E_A + E_B$ mod $E_0$, for any $E_0$.  Now if, for any $E_0$, the distribution of the modular energy $E_B$ mod $E_0$ of the ball is flat – i.e. if all values of $E_B$ mod $E_0$ between 0 and $E_0$ are equally likely – then an exchange of energy between the particle and the ball will not change the distribution of $E_B$ mod $E_0$, although it will, in general, change the distribution of $E_A$ mod $E_0$.  This one-way effect occurs if and only if the distribution of $E_B$ mod $E_0$ is flat.  But if the distribution of $E_B$ mod $E_0$ is flat, then the uncertainty in $E_B$ is at least $E_0$:

$$\Delta E_B \geq E_0 \quad .$$

We can say that a nonlocal exchange of energy of up to $E_0$ between the particle and the ball is consistent with causality, because Bob cannot detect the exchange when $\Delta E_B \geq E_0$.

For how long must the uncertainty $\Delta E_B$ be at least $E_0$?  We may reason as follows: $\Delta E_B$ must be at least $E_0$ for the whole time $T$ it takes Alice's particle to reach the piston.  This reasoning suggests an inverse relationship between the time $T$ and $\Delta E_B$, since the more energy a particle has, the less time it takes to reach the piston, and the more energy it transfers to the ball.  Indeed, let $L$ be the distance from Alice's end of the shaft to the piston and let $m_A$, $m_B$ and $p_A$, $p_B$ be the masses and momenta of the particle and the ball, respectively.  In the limit of $m_A$, $m_B$ negligibly small compared to the mass $M$ of the piston, a straightforward classical calculation shows that $E_B$ changes by $4p_A p_B/M$, if and only if Alice releases the particle. Bob can detect a nonlocal transfer of energy only if $\Delta E_B < 4 p_A p_B /M$. He cannot detect it if $\Delta E_B > 4 p_A p_B /M$. Since $p_A = m_A L /T$, we can eliminate $p_A$ to obtain

$$\Delta E_B > 4 m_A L\, p_B /MT \qquad (2)$$

as the condition for Bob not to detect the nonlocal transfer of energy.

Eq. (2) looks like an uncertainty relation for $\Delta E_B$ and $T$. But $T$ in Eq. (2) is the time Alice's particle takes to reach the piston; it is not the minimum time $\Delta t$ that Bob takes to measure $E_B$ with uncertainty $\Delta E_B$. Moreover, Bob's measurement of $E_B$ is a local measurement. The uncertainty in a local measurement depends only on local variables. The axiom of nonlocality implies that $E_B$, but not $\Delta E_B$, depends nonlocally on what Alice does. In particular, $\Delta E_B$ cannot depend on $T$ if $T$ is not a local variable. If Alice releases a particle with momentum $p_A$, then $E_B$ changes by $4 p_A p_B/M$ (according to the classical calculation), but $T$ depends on $L$ and $m_A$ as well as $p_A$. Indeed, for a given change in $E_B$, the time $T$ may be arbitrarily large. So, on the one hand, $\Delta E_B$ cannot depend on $T$. On the other hand, $E_0$ – which is the maximum nonlocal energy transfer – can certainly depend on $T$. Such a dependence is quite consistent with the axiom of nonlocality. In brief, what Eq. (2) tells us is not how $\Delta t$ depends on $\Delta E_B$, but rather how $E_0$ depends on $T$: Eq. (2) tells us that $E_0$ is inversely related to $T$.

And now, from this inverse relation, we can infer how nonlocal energy exchange could be consistent with the axiom of causality. Let $E_0 = k/T$, for some constant $k$, while $\Delta t$ is the time it takes to measure $E_B$ with uncertainty $\Delta E_B$. If the inequality $\Delta t \geq k/\Delta E_B$ holds, then $\Delta E_B \leq E_0$ implies $\Delta t \geq T$ while $\Delta t < T$ implies $\Delta E_B > E_0$, and Bob will never detect an exchange of energy before the particle reaches him. Conversely, if $\Delta t \geq k/\Delta E_B$ does not hold, we have $\Delta t\, \Delta E_B < E_0 T$ and Bob can detect the nonlocal exchange of energy by measuring $E_B$ with uncertainty $\Delta E_B < E_0$ in a time $\Delta t < T$. The axiom of causality therefore demands the inequality $\Delta t \geq k/\Delta E_B$; and for this axiom to apply consistently to nonlocal exchange of energy, $k$ must be a universal constant: Alice and Bob must not be able to circumvent the inequality $\Delta t \geq k/\Delta E_B$ by varying parameters of their experiment so as to vary $k$. Thus $k$ is a universal constant, which we can identify with Planck's constant $h$.

As noted, in quantum mechanics there is nonlocal exchange of energy. What allows quantum mechanics and causality to coexist, despite this nonlocal exchange, is the uncertainty relation for energy $E$ and time $t$ [24]. We have inverted the logical hierarchy, and from axioms of causality and nonlocality, we have derived a principle of quantum theory: the uncertainty relation for energy and time, $\Delta E\, \Delta t \geq h$.

●

Let us conclude by reviewing our progress towards a derivation of nonrelativistic quantum mechanics from the axioms of causality and nonlocality. All three attempts presented here fall short of a complete derivation. Yet there is reason for optimism. Maximal nonlocal correlations outperform quantum mechanics, but if we take a closer look, we find something quite unreasonable about them. They are so strongly correlated that, for example, Alice can actually *determine* the product of measurements of $B$ and $B'$ by choosing whether to measure $A$, or $A'$; for if she measures $A$, then $B$ and $B'$ are perfectly correlated, while if she measures $A'$, then $B$ and $B'$ are perfectly anticorrelated. Thus Alice could superluminally signal to Bob, if it were not for the (tacit) assumption that Bob cannot measure both $B$ and $B'$ but only one of them, as in quantum mechanics. In quantum mechanics, with its uncertainty relations, this assumption is natural. But in a theory with maximally nonlocal correlations, it is quite *un*natural. Bob can measure $B$ directly and also infer $B'$ indirectly from Alice's measurement. It is true that this method of measuring $B$ and $B'$ does not allow Alice to send a superluminal signal to Bob, but it defeats any uncertainty relation for $B$ and $B'$. We may still hope, therefore, that if we consider nonlocal correlations as subject to the logic of uncertainty relations, we will arrive uniquely at quantum correlations.

As for jamming, it is notable that the authors of Ref. [14] never proved the incompatibility of jamming with quantum mechanics. It just seemed obvious to them. Yet it is possible to devise a quantum thought-experiment that is equivalent to jamming. Let Alice, Bob and Jim share triplets in the GHZ state [25]:

$$|\Psi_{GHZ}\rangle = \frac{1}{\sqrt{2}} \left\{ |\uparrow\rangle_{Alice} |\uparrow\rangle_{Bob} |\uparrow\rangle_{Jim} - |\downarrow\rangle_{Alice} |\downarrow\rangle_{Bob} |\downarrow\rangle_{Jim} \right\}, \qquad (3)$$

Eq. (3) does not show the spatial wave functions of Alice, Bob and Jim, but it shows the combined state of an ensemble of three spin-1/2 atoms distributed among them. Now suppose Jim measures either the *z*-component or the *x*-component of the spin of his atom. If he measures the *z*-component of the spin, he leaves the atoms of Alice and Bob in a mixture of product states; if he measures the *x*-component of the spin, he leaves their atoms in a mixture of entangled states. Jim announces the results of his measurements, but he does not announce *what* he measured; nevertheless Alice and Bob can *deduce* what he measured if they compare the results of their measurements. They can do so, however, only in the future light cone of his announcement (which we can identify with the future light cone of *j* in Fig. 3) because they need the results of Jim's measurements, as well, to deduce what he measured. This is jamming, within quantum mechanics! Hence the question of whether quantum mechanics is the unique theory reconciling causality and action at a distance remains open after all.

These optimistic thoughts arose in the course of writing this paper, and I hope to discuss them further in a separate work.